\documentclass[aps,prl,showpacs,amsmath,twocolumn]{revtex4}
\usepackage[dvips]{graphicx}

\begin{document}

\title{Collapses and revivals of stored orbital angular momentum of light in a cold atomic ensemble}

\author{D. Moretti, D. Felinto, and J. W. R. Tabosa}
\affiliation{Departamento de F\'isica, Universidade Federal de
Pernambuco, Cidade Universit\'aria, 50670-901 Recife, PE, Brazil}

\date{\today}

\begin{abstract}
We report on the storage of orbital angular momentum of light in a
cold ensemble of cesium atoms. We employ Bragg diffraction to
retrieve the stored optical information impressed into the atomic
coherence by the incident light fields. The stored information can
be manipulated by an applied magnetic field and we were able to
observe collapses and revivals due to the rotation of the stored
atomic Zeeman coherence for times longer than 15 $\mu s$.
\end{abstract}


\maketitle

Light beams carrying orbital angular momentum (OAM) have attracted
an enormous recent interest owing to the possibility of encoding
quantum information in a multidimensional state space \cite{
Molina-Terriza07}, to their use to excite vortices in
Bose-Einstein condensates \cite{Andersen06}, as well as to a
number of others interesting applications \cite{Padgett04}. One
important family of these beams, the Laguerre-Gaussian (LG) modes
of the electromagnetic field \cite{Allen92}, possesses wave-fronts
dislocation or vortices specified by a topological charge $m$,
which sets its OAM along the propagation direction as being equal
to $m\hbar$ per photon. The coherent and nonlinear interaction of
light beams carrying OAM with atomic systems have been reported
previously using different experimental schemes \cite{Tabosa99,
Barreiro03, Barreiro06, Akamatsu03, Jiang06}. From the perspective
of quantum information processing, the use of multidimensional
state space has a promising prospect to achieve higher quantum
efficiency \cite{Bechmann-Pasquinucci00}. Indeed, entanglement
between photons with OAM and a cold atomic ensemble was already
reported in \cite{Inoue06} and more recently the generation of
twin light beams with OAM was achieved via four-wave mixing in a
hot atomic vapor \cite{Marino08}.

However, further development in this field is strongly conditioned
to our capability of reversibly store and manipulate these higher
dimensional quantum states of light into long-lived atomic
coherences. The light storage (LS) in an electromagnetically
induced transparency (EIT) medium \cite{Fleischhauer05}, which
allow us to obtain later information of a previously stored light
pulse, is a well understood phenomenon and was originally
described in terms of a mixed two component light-matter
excitation, called dark state polariton \cite{Freischhauer00}.
However, in a simpler alternative picture, LS can be described as
being due to the creation of a ground state coherence grating
which contains information on the amplitude and phase of an
optical field and which survives after the switching off of the
incident fields. To date, several experimental observations of
this phenomenon were realized in different systems
\cite{Phillips01, Liu01, Zibrov02, Mair02, Tabosa07, Moretti08}.
Recent theoretical and experimental work have also addressed the
storage of spatial structures of light beams (images) in atomic
vapors \cite{Pugatch07, Zhao08, Praveen08, Shuker08}. For
instance, a light vortex was stored in a hot vapor for hundreds of
microseconds and its robustness against diffusion demonstrated
\cite{Pugatch07}. However, to date the storage and manipulation of
superpositions of OAM states into an atomic ensemble, as well as
the characterization of the retrieved states, has not yet been
achieved.

In this work, we report the storage of superpositions of OAM
states as well as its manipulation through an applied transverse
magnetic field, which reveals the collapses and revivals of the
stored information mapped in the atomic ensemble. Since any
quantum protocols would involve necessarily arbitrary multimode
coherent superpositions of these quantum states, its storage and
manipulation, even in the classical limit, constitute a prove of
principle of such requirement. Differently from previous schemes
used to store images in thermal atomic ensembles, we employ a
delayed backward four-wave mixing (FWM) configuration which allows
us to retrieve the stored information in a different direction,
thus strongly facilitating its imagery.

For the experiment we used the Zeeman sublevels of the degenerate
two-level system associated with the cesium $D_{2}$ line cycling
transition $6S_{1/2}(F=3)\leftrightarrow 6P_{3/2}(F^{\prime}=2)$.
The cesium atoms were obtained from a magneto-optical trap (MOT)
operating in the closed transition $6S_{1/2}(F=4)\leftrightarrow
6P_{3/2}(F^{\prime }=5)$ with a recycling repumping beam resonant
with the open transition $6S_{1/2}(F=3)\leftrightarrow
6P_{3/2}(F^{\prime }=3)$. The atoms were prepared in the state
$6S_{1/2}(F=3)$ by switching off the repumping beam for a period
of about 1 ms to allow optical pumping by the trapping beams via
non-resonant excitation to the excited state $F^{\prime }= 4$. The
MOT quadrupole magnetic field is also switched off during these
optical pumping period. Typically, the measured optical density of
the sample of cold atoms in the $F=3$ ground state is
approximately equal to 3 for appropriate MOT parameters. Three
pairs of Helmholtz coils are used to compensate for residual
magnetic fields. In Fig. 1 (a), (b) we show a generic $\Lambda$
three-level system consisting of two degenerate ground states and
one excited state belonging to the corresponding hyperfine Zeeman
manifold, coupled with the incident beams at two different
instants of time, according to the time sequence showed in Fig.
1-(c) and the beam geometry of Fig. 1 (d). As depicted in these
figures, the two incident writing beams, labelled as $W$ and
$W^{\prime}$, have opposite circular polarization and are incident
on the MOT forming a small angle $\theta \approx 3^{0}$. These
beams will therefore excite a ground-state coherence grating into
the atomic ensemble. The induced coherence grating is probed by a
reading beam $R$, which is counter-propagating to the writing beam
$W$ and has a circular polarization opposite to that beam. In the
continuous wave (cw) excitation of the ensemble, this corresponds
to the well-known backward FWM configuration \cite{Lezama01}. For
the time sequence shown in Fig. 1(c), the retrieved signal
corresponds to the Bragg diffraction of beam R into the stored
coherence grating and as shown below is counter-propagating to the
write beam $W^{\prime}$.

\vspace{-1.2 cm}
\begin{figure}[htb]
\includegraphics[angle=0,width=8.5cm]{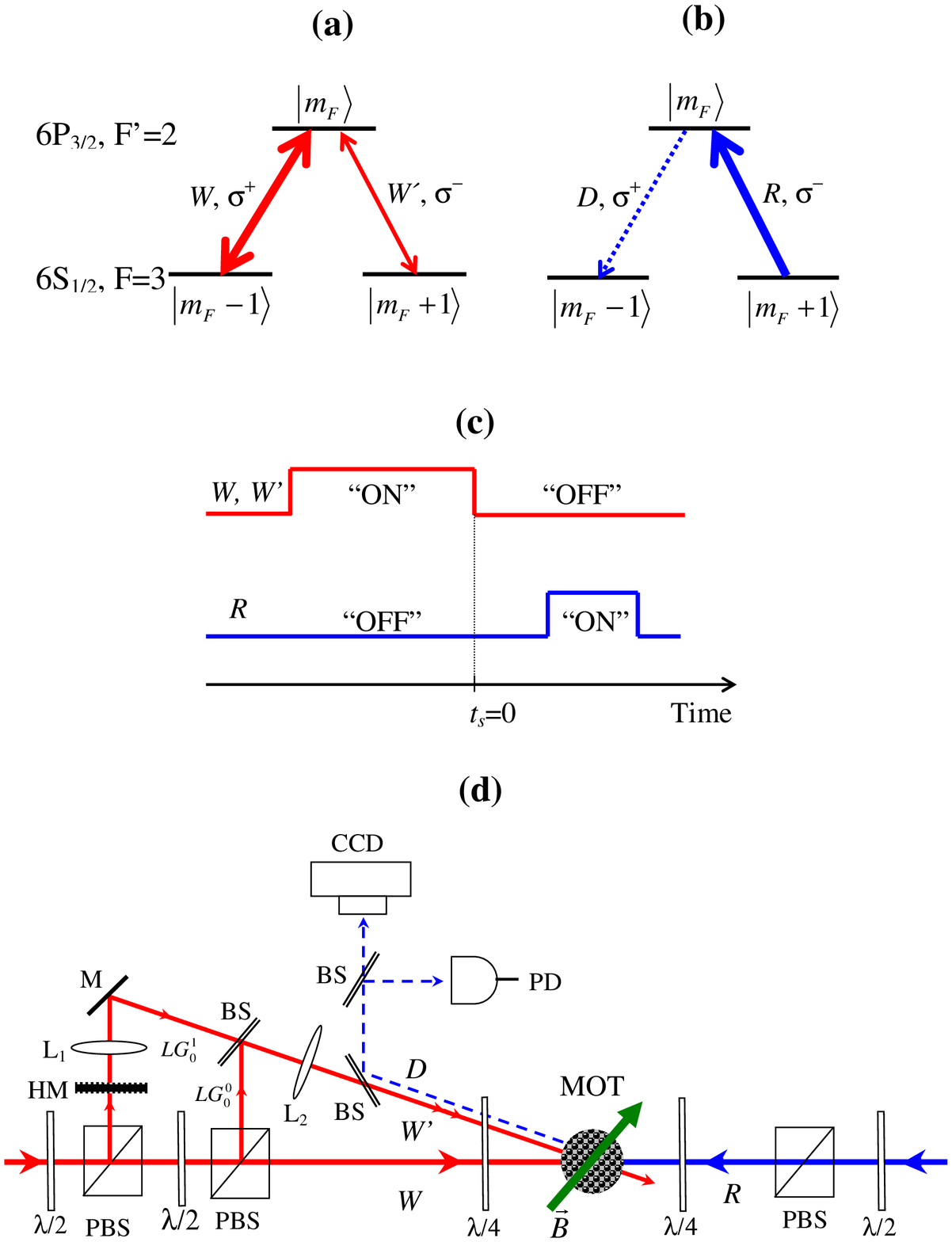}
\vspace{-1.0cm} \caption{(a) Generic Zeeman three-level system,
showing the coupling with the grating writing beams $W$ and
$W^{\prime}$. Here the beam $W$ has a much higher intensity than
the beam $W^{\prime}$, so the atoms are optically pumped into the
highest $m_{F}$ level and this level scheme would correspond to
$m_{F}=2$. (b) The coupling of the reading beam $R$ with the
coherently prepared three-level atom, generating the diffracted
beam $D$. (c) The switching time sequence for the writing and
reading beams. (d) Simplified experimental scheme. All the beams
are provided by a diode laser locked to the $F=3\leftrightarrow
F^{\prime}=2$ transition, and are intensity modulated using
acousto-optical modulators not shown in the scheme. $\vec{B}$ :
Magnetic field; PBS: Polarizing beam splitter; BS: Beam splitter;
HM: Holographic mask; CCD: Camera; PD: Photodetector; L: Lens.}
\end{figure}

In Ref.~\onlinecite{Moretti08} we showed that the diffracted field
$D$ generated by the atomic ensemble is given by the coherent sum
of contributions from all excited atomic dipoles, with such
dipoles obtained from the expression for the slowly varying
coherences
\begin{equation}
\sigma_{2,1a}(\vec{r},t) = \frac{g_R(t) e^{-\gamma
t_s}\Omega_R(\vec{r})\Omega_W(\vec{r})\Omega_{W^{\prime}}^*(\vec{r})}{|\Omega_W(\vec{r})|^2
+ |\Omega_{W^{\prime}}(\vec{r})|^2}\,,
\end{equation}
where, according to Fig. 1 (a) and (b) we labelled $2$ and $1a$
the excited state $|m_F\rangle$ and the ground state $|m_F -
1\rangle$, respectively. In this equation, $\Omega_i(\vec{r})$
represents the Rabi frequency associated with beam $i$ in an atom
at position $\vec{r}$, being given then by
$\Omega_i(\vec{r})=\mu_{a,b} {\cal E}_i(\vec{r})e^{i
\vec{k}_i\cdot \vec{r}}/\hbar$, with $\mu_{a,b}$ the dipole moment
between the levels $a$ and $b$ coupled by $\Omega_i$, ${\cal
E}_i(\vec{r})$ the electric field envelope of beam $i$, and
$\vec{k}_i$ its wave-vector. The term $e^{-\gamma t_s}$ is a
consequence of the effective decay of the ground state coherence
due to a residual inhomogeneous magnetic field, with $\gamma$ the
decay rate and $t_s$ the ``storage time''. Finally, $g_R(t)$ is a
function describing the shape of the light pulse in mode $D$
retrieved by the read pulse $R$:
\begin{equation}
g_R(t) = \frac{e^{-\gamma_1 t}\sinh (\gamma_2 t)}{\gamma_2} \,,
\end{equation}
with $\gamma_1 = \Gamma_{22}/2 + \gamma$, $\gamma_2 =
\sqrt{(\Gamma_{22}/2-\gamma)^2-4|\Omega_R|^2}/2$, and
$\Gamma_{22}$ is the spontaneous decay rate of level 2. In the
above equations $t=0$ corresponds to the start of the reading
pulse. The retrieved pulse shape and its saturation and decay
dynamics were extensively studied in Ref.~\onlinecite{Moretti08}.

In our present experiment, fields $W$ and $R$ can be well
approximated  by plane waves exciting the atomic ensemble. On the
other hand, since the waist of beam $W^{\prime}$ is considerably
smaller than the cloud of atoms, the transverse profile of beam
$W^{\prime}$ is what actually defines the excited-ensemble shape.
Taking this into account, the fact that $\vec{k}_W = -\vec{k}_R$,
and that the intensity of $W$ is much larger than the one of
$W^{\prime}$, the expression for the excited coherence responsible
for the retrieved field $D$ can be well approximated by
\begin{equation}
\sigma_{2,1a}(\vec{r},t) = A \, g_R(t) e^{-\gamma t_s}{\cal
E}_{W^{\prime}}^*(\vec{r})e^{-i\vec{k}_{W^{\prime}}\cdot
\vec{r}}\,, \label{coherence}
\end{equation}
with $A$ a constant in position and time, depending only on the
pumping-beams intensities and the dipole moments of the involved
transitions. Equation~\eqref{coherence} explicitly indicates that,
in our experimental conditions, the $D$ field propagates in the
direction ($-\vec{k}_{W^{\prime}}$) opposite to $W^{\prime}$, and
have a transverse beam profile given by ${\cal
E}_{W^{\prime}}^*(\vec{r})$. As we will show, for a
Laguerre-Gaussian mode in ${\cal E}_{W^{\prime}}(\vec{r})$, this
implies that the OAM of the retrieved beam will be inverted with
respected to the $W^{\prime}$ beam. Equation~\eqref{coherence}
also shows that the retrieved pulse amplitude decays with a rate
determined by the ground-state-coherence lifetime.

We consider now explicitly the case where the incident writing
beam $W^{\prime}$ contains Laguere-Gaussian modes with topological
charge $m$, i.e., $LG_{0}^{m},$ described, in polar coordinates
$(r,\phi, z)$ in the plane $z=0$, by a field amplitude ${\cal
E}_{W^{\prime}} \propto
(\sqrt{2}r/\omega_{0})^{m}\exp(-r^2/\omega_{0}^2)e^{-im\phi}$,
where $\omega_0$ is the minimum beam waist. As shown in Fig. 1
(d), the writing beam $W^{\prime}$ can be composed either by a
Laguerre-Gaussian mode with a topological charge $m=1$,
i.e.,$LG_{0}^{1}$, generated by the holographic mask $HM$ and the
lens $L_{1}$, or by a zero charge Laguerre-Gaussian mode, or by a
superposition of these two modes. These two beams, have
approximately the same power of $20 \mu W$ and are focused by the
lens $L_{2}$ into the MOT with a diameter much smaller than the
size of the trapped atomic cloud ($\approx 2~mm$). The writing
beam $W$ is a simple Gaussian mode ($LG_{0}^{0}$), has a power of
$\approx 2.0~mW$ and a diameter of the order of the trapped cloud.
A pair of acousto-optical modulators (AOM) is used to scan the
common frequency of the writing beams $W$ and $W^{\prime}$ around
the atomic resonance, as well as to allow their fast switching.
The reading beam $R$, has Gaussian intensity profile, power, and
diameter of the same order of beam $W$, passing also through
another pair of AOM's.

\vspace{-1.2cm}
\begin{figure}[htb]
\hspace*{-0.0cm}\includegraphics[angle=0,width=7.5cm]{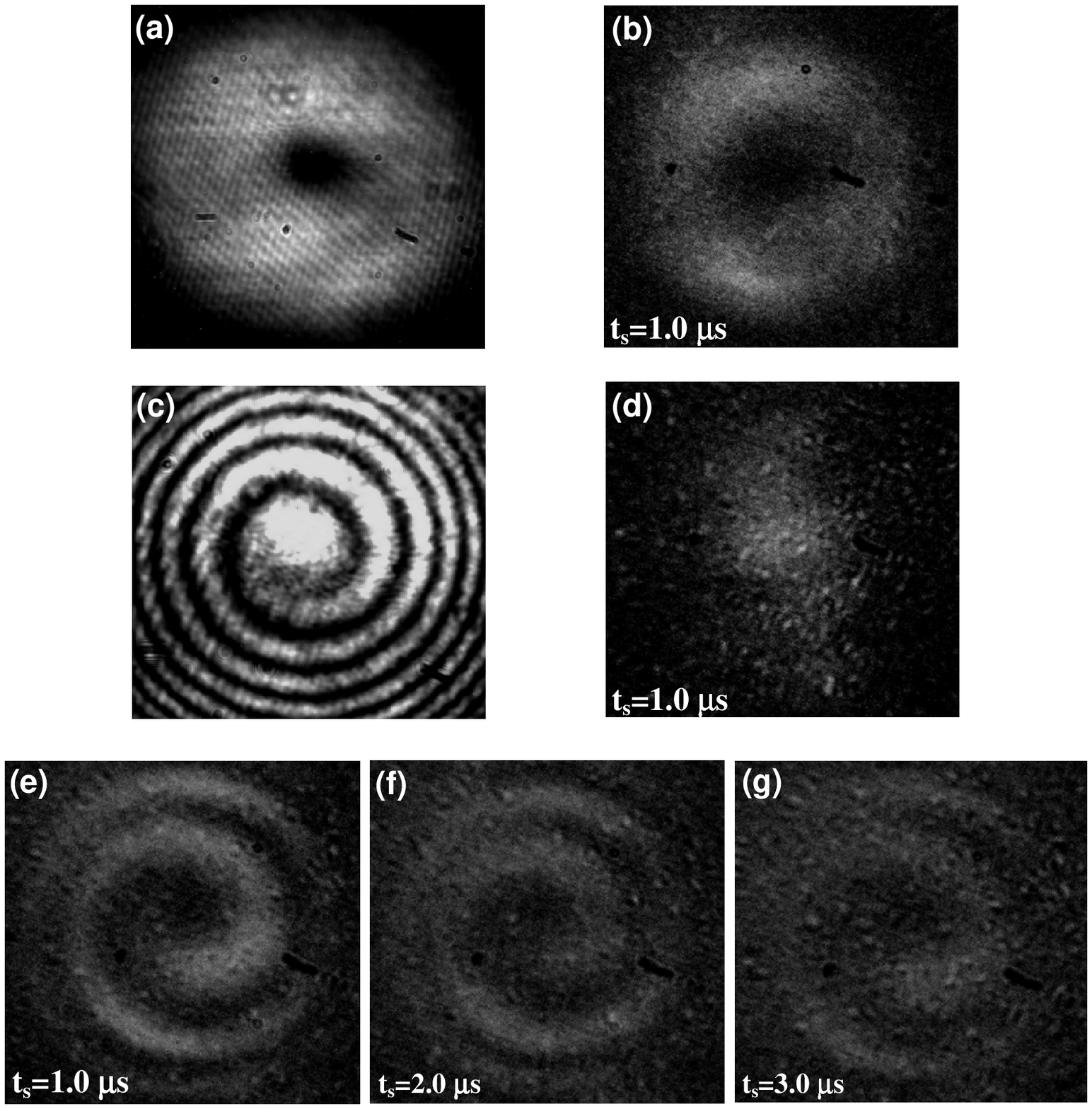}
\vspace{-2.7 cm} \caption{(a) The incident writing beam intensity
profile for the case $W^{\prime}=LG_{0}^{1}$; (b) and (d) The
retrieved beam associated respectively with an incident writing
beam with modes $W^{\prime}=LG_{0}^{1}$ and
$W^{\prime}=LG_{0}^{0}$; (c) Incident writing beam profile
corresponding to a beam superposition
$W^{\prime}=LG_{0}^{1}+LG_{0}^{0}$. This image is obtained by
retro-reflecting the incident superposition right after the MOT.
(e), (f) and (g) the retrieved signal corresponding to the
incident superposition shown in (c) for the different storage
times indicated in the figure. All this set of measurements is
obtained without magnetic field.}
\end{figure}

Under the EIT resonant condition, we have recorded the image of
the retrieved beam $D$ for different incident $W^{\prime}$ modes:
In Fig. 2 (a) and (b) we show respectively the image of the
incident $W^{\prime}=LG_{0}^{1}$ mode and the image of the
corresponding retrieved beam. In order to analyze the topological
charge of the retrieved beam, we consider as a reference wave the
diffracted beam associated with a incident Gaussian beam
$W^{\prime}=LG_{0}^{0}$, whose retrieved profile is shown in Fig.
2 (d). In Fig. 2 (c) we show the image for the incident
superposition $W^{\prime}=LG_{0}^{1}+LG_{0}^{0}$, obtained by
retro-reflecting it with a mirror. In Fig. 2 (e), (f) and (g) we
show the corresponding retrieved images obtained for the different
storage times indicated in each image. The retrieved signal
intensity, as measured simultaneously with the photodetector,
decays exponentially with a time constant of approximately $3\mu
s$, which we attribute mainly to non-compensated magnetic field
gradients~\cite{Moretti08}. Clearly, these images reveal that
incident and the retrieved beams have the same single topological
charge. However, in order to determine its corresponding sign, we
carefully have to take into account that the sense of rotation of
the spiral in the interference pattern (which is determined by the
sign of $m$) changes upon reflection on a mirror and after passing
through a focus \cite{Barreiro03}. This leads to two consecutive
reversions in the sense of the spiral associated with the imagery
of the incident $W^{\prime}$ superposition. According to the
recorded interference patterns shown in Fig. 2 (c) and (e)-(g), we
can therefore conclude that the incident and the retrieved beams
have the same sign for its topological charge. Moreover, as these
two beams propagates in opposite directions they should carry
opposite OAM. These observations not only demonstrates that
superpositions of OAM states can be stored into the atomic
ensemble, but also the reversible transfer of OAM between light
and atoms.

In another series of experiments, we applied an external $dc$
magnetic field of magnitude $B\approx 0.6$ G, nearly orthogonal to
the plane defined by the incident beams, and therefore
perpendicular to the quantization axis, defined along the beams
propagation direction. In this case we observe a series of
collapses and revivals of the stored Zeeman coherence as can be
seen in Fig. 3, which shows the amplitude of the retrieved signal
associated with the input writing beam
$W^{\prime}=LG_{0}^{1}+LG_{0}^{0}$ for different storage times. It
is worth noticing that a similar collapses and revivals
observation has been reported before and was interpreted as being
due to the Larmor precession of the collective spin excitation
around the applied magnetic field \cite{Matsukevich06}.
Differently from the results presented in Fig. 2, we have observed
approximately a four-fold increase in the decay time of the stored
coherence grating in the presence of the applied magnetic field.
Probably this is associated to a reduced effect of the magnetic
field inhomogeneity due to the much stronger applied uniform
magnetic field.

For the applied magnetic field, the associated Larmor period
$T_{L}=2\pi/\Omega_{L}$, where $\Omega_{L} = g_{F}\mu_{B}B/\hbar$
is the Larmor frequency, $g_{F}$ the Lande factor for the lower
ground state $F=3$ and $\mu_{B}$ the Bohr magneton, is  $T_{L}
\approx 5 \mu s$ \cite {Steck}. Thus, revivals of the stored
coherence should occur for integer multiples of the Larmor period.
Furthermore, at half the Larmor period the Zeeman coherence and
populations are just reversed and therefore we should also expect
partial revivals of the retrieved signal for odd multiples of half
the Larmor period. Theses considerations are completely consistent
with the series of revivals observed in Fig. 3. We have also
recorded \linebreak

\vspace{-2.5cm}
\begin{figure}[tbh]
\hspace*{-1.5cm}\includegraphics[angle=0,width=12cm]{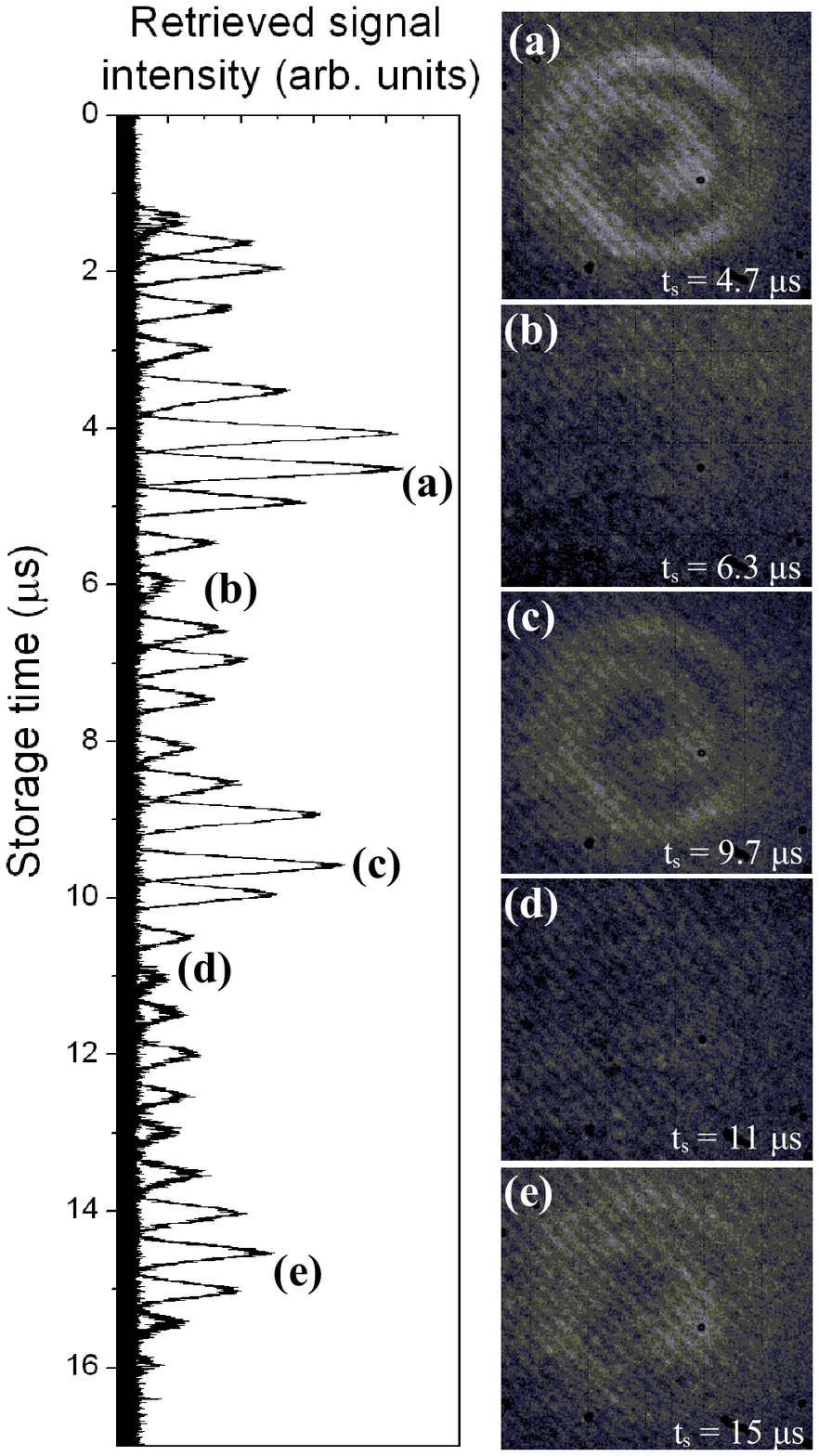}
\vspace{-3.5cm} \caption{The amplitude of the retrieved signal
associated with a input writing beam
$W^{\prime}=LG_{0}^{1}+LG_{0}^{0}$ for different storage times
when a transversal $dc$ magnetic field is applied.  Right column:
Retrieved signal intensity profile for the times indicated in the
images (a), (c), (e), which correspond to times associated with
the first three principal branches of revival peaks; (b) and (d):
Frames taken for times corresponding to the indicated collapses
times. Note that no images can be seen in these frames.}
\end{figure}

\noindent the images of the retrieved beam for different storage
times in the presence of the magnetic field, as shown in the right
column of Fig. 3. As indicated, the images (a), (c) and (e) where
recorded for storage times corresponding respectively to the first
three principal branch of revivals, while images (b) and (d) where
recorded for times in the collapse regions, as also indicated in
the figure. These results clearly show that the stored collective
ground state coherence, containing the information on the
superposition of OAM states of the incident beam, can be
manipulated by the external magnetic field. To our knowledge, this
constitute the first demonstration of such manipulation of OAM
stored in atomic coherences.

In summary, we have experimentally demonstrated the storage of OAM
of light in a cold ensemble of cesium atoms, as well as the
reversible exchange of OAM between light and atoms. The
manipulation of the stored OAM was also demonstrated via Larmor
precession of a spatially distributed ground state coherence in an
external magnetic field. In particular, collapses and revivals of
the stored coherence were directly observed and reveal the control
of a collective light-matter excitation. We believe our results
fill up one of the gaps challenging the implementation of quantum
information processing encoded in a multidimensional state space.

We gratefully acknowledge fruitful discussion with A. Lezama. This
work was supported by the Brazilian Agencies CNPq/PRONEX,
CNPq/Inst. Mil\^{e}nio and FINEP.

\end{document}